\documentclass[aps,prl,superscriptaddress,twocolumn]{revtex4-2}

\usepackage{graphicx}

\begin{document}


\title{Supercontinuum amplification by Kerr instability}


\author{Sagnik Ghosh}
\affiliation{Dept. of Electronics and Communication Engineering, Manipal Institute of Technology, Udupi-Karkala Rd, Eshwar Nagar, Manipal, Karnataka 576104 India}
\author{Nathan G. Drouillard}
\affiliation{Dept. of Physics, University of Windsor, Windsor ON N9B 3P4 Canada}
\author{TJ Hammond}
\email{thammond@uwindsor.ca}
\affiliation{Dept. of Physics, University of Windsor, Windsor ON N9B 3P4 Canada}


\date{\today}

\begin{abstract}
The versatility of optical parametric amplifiers makes them excellent sources for ultrashort visible and near-infrared pulses that drive strong-field physics experiments. We extend four-wave optical parametric amplification to the strong-field regime, known as Kerr instability amplification, altering the non-collinear conditions for optimum amplification. Experimentally, we amplify spectra that span nearly an octave without spatial chirp. Additionally, we investigate the conditions necessary for amplifying single-cycle pulses in the near infrared; this finding is particularly relevant for generating isolated attosecond pulses from narrow bandgap semiconductors.
\end{abstract}


\maketitle

\section{Introduction}

Next-generation coherent light sources use nonlinear parametric amplification to generate intense pulses from the visible to the infrared (IR) \cite{FattahiOptica2014}. These versatile ultrafast sources have a broad range of applications including spectroscopy \cite{BiegertIEEE2012}, high harmonic generation \cite{ShinerNatPhys2011}, and attosecond (1 as = 10$^{-18}$ s) science \cite{VampaNature2015, HammondNatPhot2017, ZinchenkoScience2021}. These optical parametric amplifiers (OPAs) are often pumped by intense femtosecond (1 fs = 10$^{-15}$ s) pulses, generating widely tunable and broad bandwidth coherent pulses with improved pulse contrast \cite{KobayashiJPB2012, WagnerAPB2014, DorrerJOpt2015, SunApplOpt2021}. The non-inversion symmetry required of the $\chi^{(2)}$ optical nonlinearity and phase matching conditions limit the selection of possible crystals \cite{ManzoniJOpt2016}. However, non-collinear geometries circumvent phase matching limitations, increasing the amplified bandwidth \cite{CerulloRSI2003}.

The next order nonlinearity, $\chi^{(3)}$, has also been investigated to amplify visible and near-IR femtosecond pulses \cite{HansrydIEEE2002, DubietisOE2007, ValtnaOL2008, HuangOLT2022}. In this case, phase matching is achieved when $2 \mathbf{k_p} = \mathbf{k_s} + \mathbf{k_i}$, where $\mathbf{k_{p,s,i}}$ are the wave-vectors of the pump, signal, and idler, respectively, which is only achieved in a non-collinear geometry. Amplification of up to two orders of magnitude was observed, with a bandwidth that could span up to 75 nm in the visible \cite{DubietisLC2008}. When the amplified signal reaches $\sim1\%$ of the pump intensity, saturation limits further amplification, resulting in a non-degenerated cascaded four-wave mixing (NDC FWM).

NDC FWM creates a series of high-order beamlets spanning multiple octaves \cite{CrespoOL2000}. Although these beamlets propagate in unique directions, spatial dispersion compensation has compressed these pulses to the few-cycle regime \cite{WeigandPRA2009}. Furthermore, these beamlets have been used as a tuneable probe for fs stimulated Raman spectroscopy \cite{ZhuAPL2013, ZhuAPL2014, DietzeCPC2016}.

In more extreme case, where the amplified seed remains below saturation and the pump intensity is near the damage threshold of the $\chi^{(3)}$ material, Kerr instability amplification (KIA) has been proposed to amplify far-IR and THz pulses from a near-IR pump \cite{NesrallahOptica2018}. It demonstrated over three orders of magnitude amplification with a Gaussian beam profile \cite{VampaScience2018}. One of the authors previously reported experimental amplification of widely tuneable but spectrally limited fs pulses using Kerr instability in Y$_3$Al$_5$O$_{12}$ (YAG) \cite{VampaScience2018} and magnesium oxide (MgO) \cite{JACOL2021}. In the visible and the near-IR, gain as high as 18/mm was measured. 

In this current work, we measure and simulate the amplification of a supercontinuum from KIA in MgO. Although saturation and geometric effects limit the experimentally measured amplification to two orders of magnitude, our simulations and calculations show that up to five orders of magnitude amplification are realizable.

\section{Angle-dependent Amplification}

The experimental setup is shown in Fig. 1(a). The 785 nm Ti:Sapphire laser output 1 mJ, 110 fs pulses at 1 kHz. A CaF$_2$ window (not shown) split $\approx 8\%$ of the $s$ polarized beam to create the seed; the half wave plate (HWP) and polarizer (Pol.) controlled the seed power to 1~\textmu J, while the iris controlled the beam profile to optimize the supercontinuum generated through the focus ($f_1 = 10$~cm) of 5 mm sapphire. The supercontinuum spectrum spanned from 450 to 1000~nm at 50~dB, measured using four different wavelength selecting filters to optimize the signal across the spectrum (350 - 650~nm, 600 - 750~nm, 800 - 900~nm, and $>850$~nm, where the overlap ensured continuity in the spectrum), measured with an OceanOptics Flame-S spectrometer. The collimating mirror ($f_c = 10$~cm) refocused the seed beam onto the MgO, with $\approx 5.2\times$ magnification; $M_1$ steered the seed beam, where the external seed-pump angle was $\theta$.

\begin{figure}[h]
\includegraphics[width=1\columnwidth]{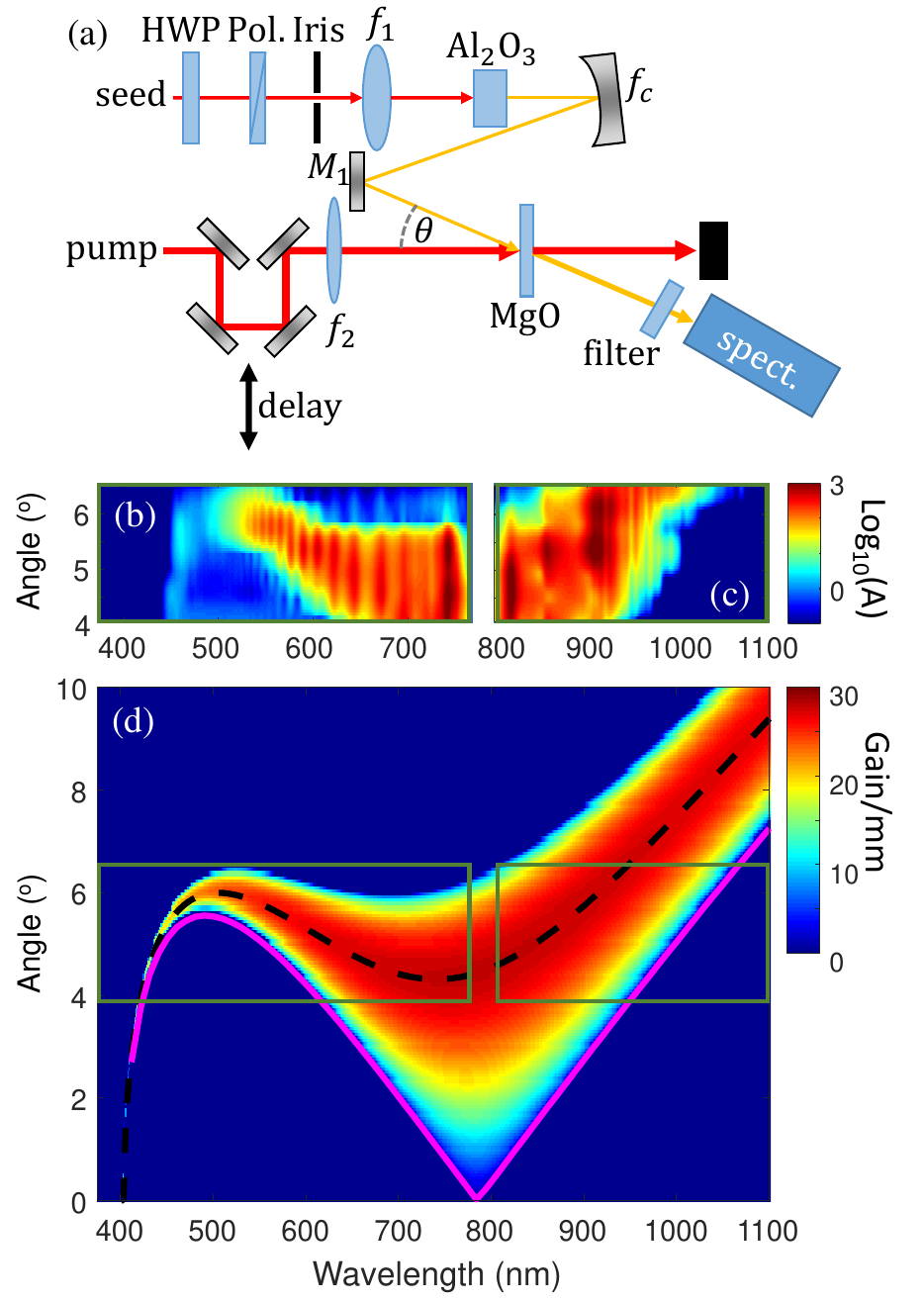}
\caption{(a) The setup for supercontinuum amplification. See text for details. Spectral amplification (log colors) measured in the visible (b) and near IR (c) as a function of seed-pump angle. (d) Kerr instability amplification calculation of the gain as a function of relative seed-pump angle and wavelength for 785 nm pump at $I_0 = 1.5\times10^{13}$ W/cm$^2$. The pink line is the phase matching condition for FWM, while the black dashed line is maximum gain in KIA. The two green boxes are for the experimentally measured regions, shown in (b) and (c).}
\end{figure}

A computerized delay stage controlled the pump timing; $f_2 = 50$~cm loosely focused ($w_0 \approx 100$~\textmu m) the pump (pulse energy 420 \textmu J) onto the 0.5~mm MgO crystal. The peak intensity was estimated to be $1.3 \times 10^{13}$~W/cm$^2$. Strong-field experiments have shown that MgO can withstand pump intensities up to $I_p = 1.5 \times 10^{13}$~W/cm$^2$ without damage \cite{YouNatPhys2016, KorobenkoOptExp2019, KoJPB2020, JACOL2021}. Wavelength and neutral density filters selected the spectrum and intensity for the spectrometer (spect.); the same filters for measuring the seed spectrum were used for the amplified spectra.

We measured amplification in the (b) visible, and (c) IR. The relative seed-pump angles were 4.9 to 6.5$^\circ$ with 0.4$^\circ$ steps. Because of the supercontinuum chirp, the amplified spetra were for a delay that showed the maximum bandwidth (see Fig. 2(b) for further details). We observed an angle dependence on the amplified bandwidth; at 4.1$^\circ$, with amplification occurring from 650 to 950~nm. As we increased the relative angle, the amplified bandwidth maximizes near 5$^\circ$, spanning from 550 to nearly 1000 nm, or nearly an octave. In this region, the transform limited pulse duration was 5~fs, or two-cycles with a central wavelength of 724~nm. Further increasing the angle led to a bifurcation in the amplified spectrum, where the visible portion around 530~nm was amplified, although with decreased gain. At the maximum measured angle, we observed large amplification beyond 1000~nm, measured with an OceanOptics NIRQuest spectrometer. 

Saturation limited the maximum amplification to roughly two orders of magnitude. Because we aimed to preserve the generated supercontinuum character for the amplification measurement, we avoided using transmission ND filters to decrease the seed power. We replaced the steering mirror $M_1$ with BK7 glass to reduce the seed power into the MgO; the seed power was $\approx  70$~nJ. The amplification process saturates when the amplified pulse is $\sim 1\%$ of the pump intensity \cite{JACOL2021}. Above saturation, we observed multiple spatially-separated beams spanning the UV to the IR, as predicted in NDC FWM \cite{WeigandApplSci2015}.

\begin{figure}[b]
\includegraphics[width=1\columnwidth]{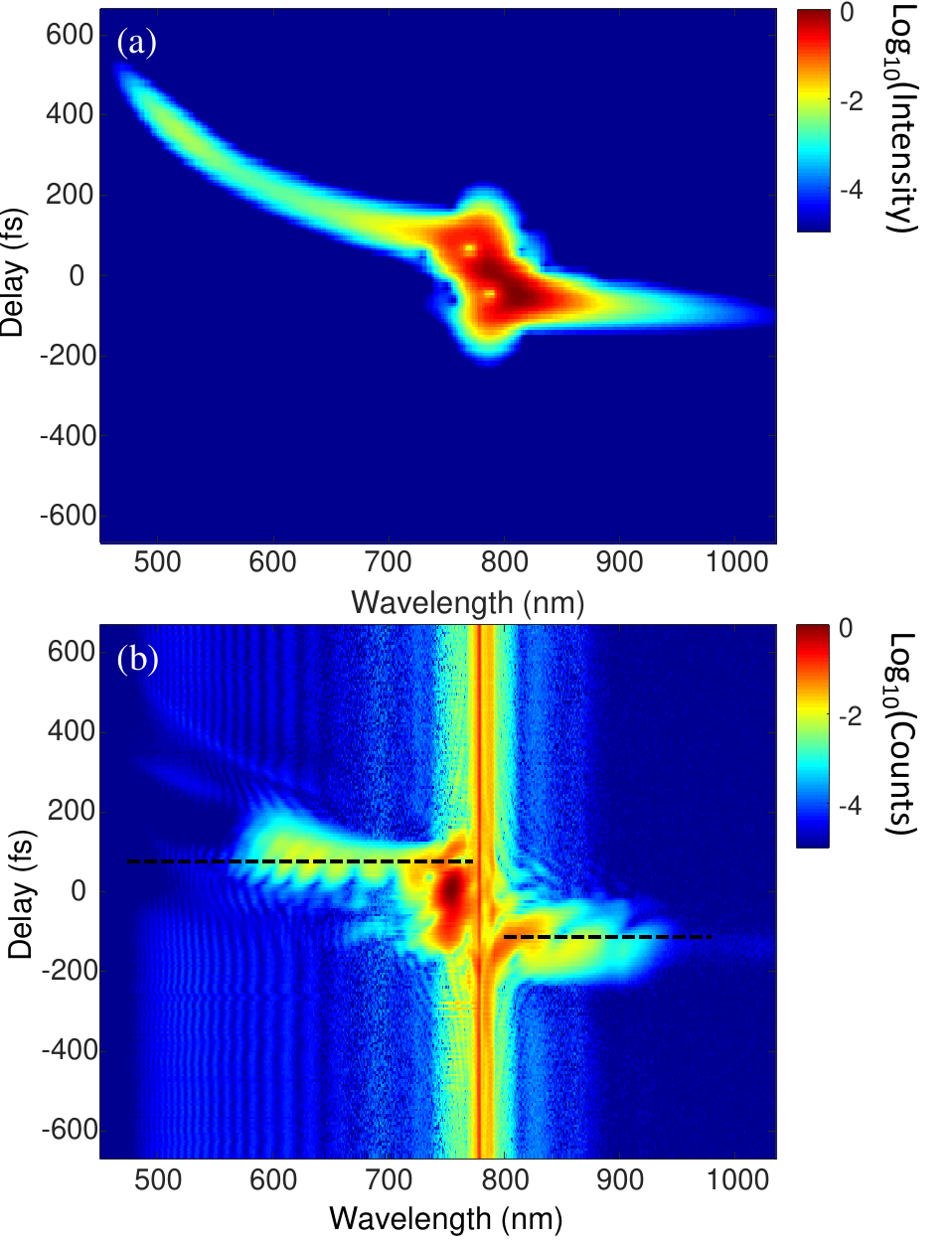}
\caption{(a) Temporal evolution of simulated supercontinuum generation matching experimental conditions. (b) Spectral amplification measurement of the supercontinuum seed at 4.9$^\circ$ relative seed-pump angle as a function of pump delay. The amplification process maintains the seed temporal character. Black dashed lines are the regions used to generate the 4.9$^\circ$ contributions in Fig. 1 (b) and (c).}
\end{figure}

The gain predicted by KIA is shown in Fig. 1(d), where the experimentally measured regions are shown in the green rectangles for comparison. In the limit of low pump intensity, the angular dependence of the maximum gain predicted by KIA (black dash) is identical to the phase matching condition of FWM (pink solid), but at this high intensity (high gain) case there is an obvious difference. Near the pump wavelength and longer, the angle for maximum gain KIA is considerably larger than that predicted by FWM. KIA predicts a region of large gain over a broad bandwidth with a small angular dependence, which leads to broadband amplification. The false colour scale is the gain predicted by KIA, where near 5$^\circ$ it predicts a region of near uniform amplification spanning 550 to 1000~nm, in agreement with our experiment. See Supplemental Information for KIA calculation details.

We simulated the temporal profile of the generated supercontiuum in Fig. 2(a), replicating the experimental parameters. A 120~fs seed pulse focused to 25~\textmu m at $4.5 \times 10^{15}$~W/cm$^2$ peak intensity propagated 5 in mm sapphire (see Supplemental Information for simulation details). We took the windowed Fourier transforms (Gaussian window width 40~fs) of the output pulse to show the temporal character. As the pulse propagated through the focus, the pulse split into a leading IR pulse and a tailing visible pulse. After the focus, the visible portion of the pulse developed a significant chirp due to the material dispersion. 

The measured spectrally dependent amplification as a function of pump delay is shown in Fig. 2(b) for 4.9$^\circ$. Near -100~fs, the IR portion was amplified, while at +100~fs the visible portion was ampified, with an increasing delay for shorter wavelengths. The temporal profile of the amplification is in good agreement with the simulated supercontinuum, demonstrating that KIA maintains the temporal profile of the seed. Black dash lines represent the regions contributing to Fig. 1(b) and (c).

\begin{figure}[h]
\includegraphics[width=1\columnwidth]{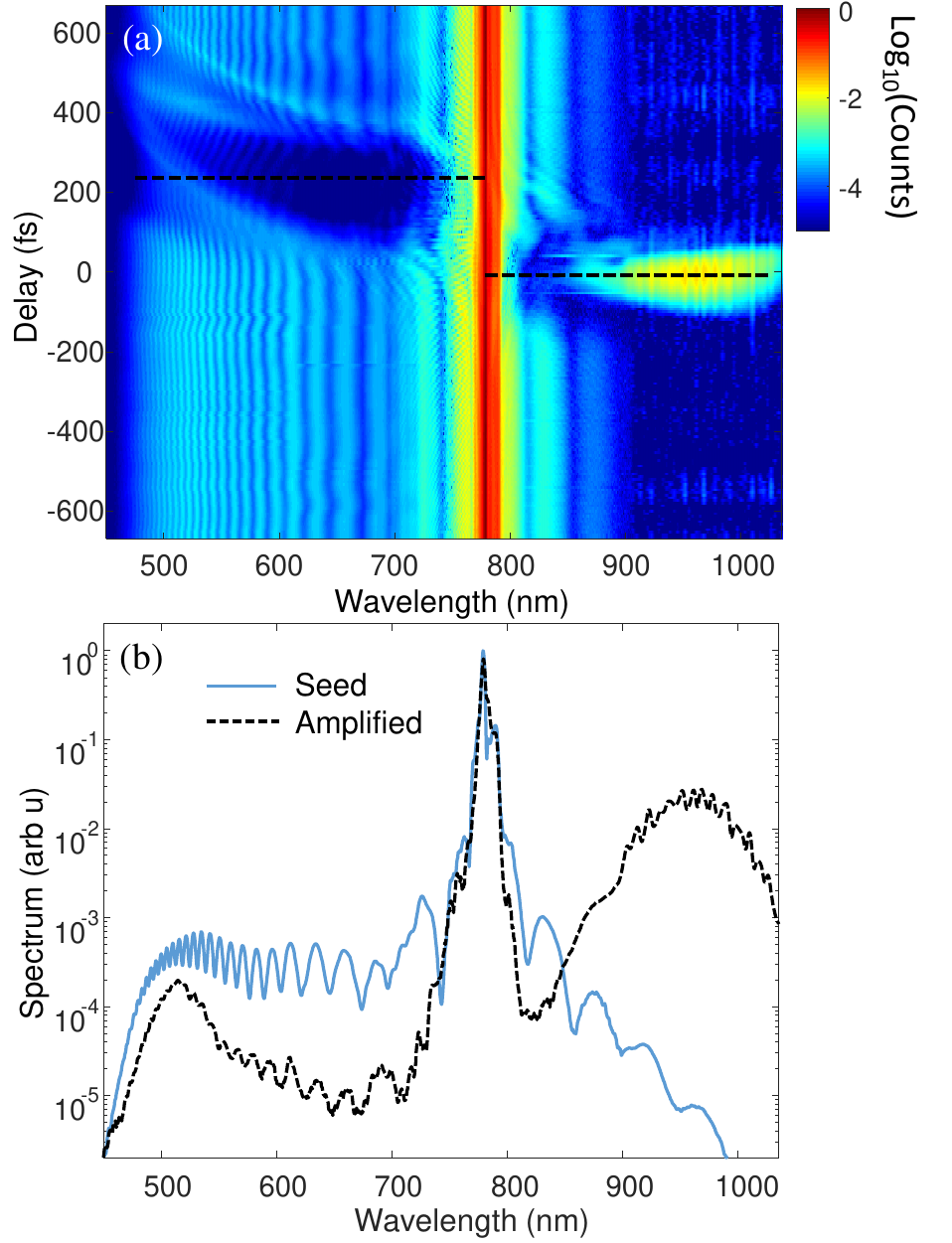}
\caption{Spectral amplification measurement of the supercontinuum seed at 6.5$^\circ$ relative seed-pump angle. (a) Amplification occurs only for the IR side of the seed. Black dashed lines are lineouts to generate spectrum in (b). (b) Supercontinuum seed (blue) and amplified (black dashed) spectra. The wavelength region $> 900$~nm was taken with an IR spectrometer.}
\end{figure}

Increasing the angle to 6.5$^\circ$, we measured the resulting amplified spectra as a function of pump delay in Fig. 3(a). We observed amplification factors $\sim$ three orders of magnitude in the IR. The increased amplification factor was due to the low seed power in this spectral region. Conversely, in the visible, the Kerr instability conditions were not satisfied, which led to a decreased visible signal. Interestingly, the temporal character of the seed was maintained even in this regime. Black dashed lines are the time delays for the lineouts in (b). We show the supercontinuum spectrum (blue) spanning from the visible to the IR that seeds the amplifier. At 6.5$^\circ$, we observe amplification around 950~nm, in agreement with the KIA calculation. However, the visible portion decreases by more than an order of magnitude, which is not expected from KIA.

\section{Broadband Amplification}

Although it has been reported that NDC FWM generates broadband pulses \cite{WeigandPRA2009}, KIA can amplify few-cycle pulses without spatial chirp, as shown in Fig. 4. In these simulations and experiment, we observed the output at 4.9$^\circ$. The KIA calculation (black) predicts a broadband amplification of nearly six orders of magnitude, with a full width at half maximum (FWHM) spanning from 600 to 900~nm. This amplification corresponds to a gain of 27/mm, with the transform limited pulse duration being 4.7~fs and a central wavelength of 714~nm, supporting a two-cycle pulse. We simulated the nonlinear propagation of the FWM process by seeding with a broadband seed (red). Further simulation details are found in the Supplemental Information. The simulation used the experimental parameters, but the pump pulse character was modified to 800~fs in duration and a minimum waist of $w_p = 250$~\textmu m to avoid self-phase modulation and self-focusing effects of the pump (the plane wave long-pump limit of KIA). Although the simulated amplified spectrum matches the KIA results, FWM created the idler beam within the first 100~\textmu m of the crystal, which decreased the signal amplification and was not accounted for in KIA theory.

\begin{figure}[h]
\includegraphics[width=1\columnwidth]{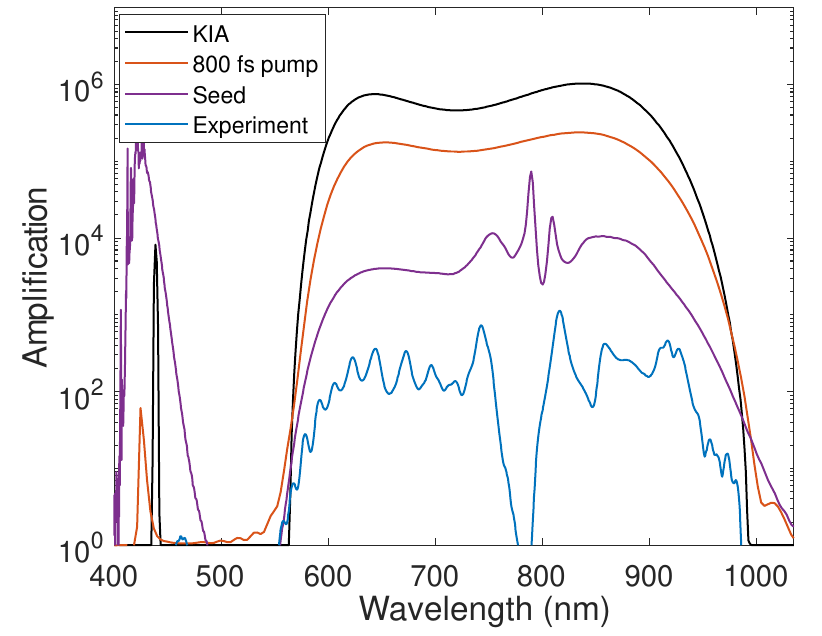}
\caption{Simulation and experimental amplification at 4.9$^\circ$. The KIA calculation (black) agrees well with the 2D simulation (red); the amplification difference comes from KIA not accounting for the creation of an idler field. The spatial profile reduced the amplification of the transform limited supercontinuum seed pulse when the seed and pump beam waists were the same (purple); the large gain in the blue end of the spectrum is because there was negligible seed spectrum in this region. The bandwidth of the experimentally measured amplified spectrum is in good agreement with theory.}
\end{figure}

We compare this broadband amplification to the amplification of the simulated seed previously discussed (purple). In this case, the seed duration transform limit was 10 fs, while the seed waist matched the pump waist, $w_s = w_p$, and the pump pulse duration was 100~fs, matching experimental conditions. The seed profile and the shorter pump duration decreased the amplification relative to the maximum possible amount. We also show the experimentally realized amplification (blue). The bandwidth of the amplified spectrum agrees well with theory, but the maximum amplification is significantly decreased from the theoretical value because of saturation. To measure the amplification factor, we smoothed the seed spectrum to reduce the oscillation amplitude. We note that there was little amplification near the seed central wavelength in this configuration.

\begin{figure}[h]
\includegraphics[width=1\columnwidth]{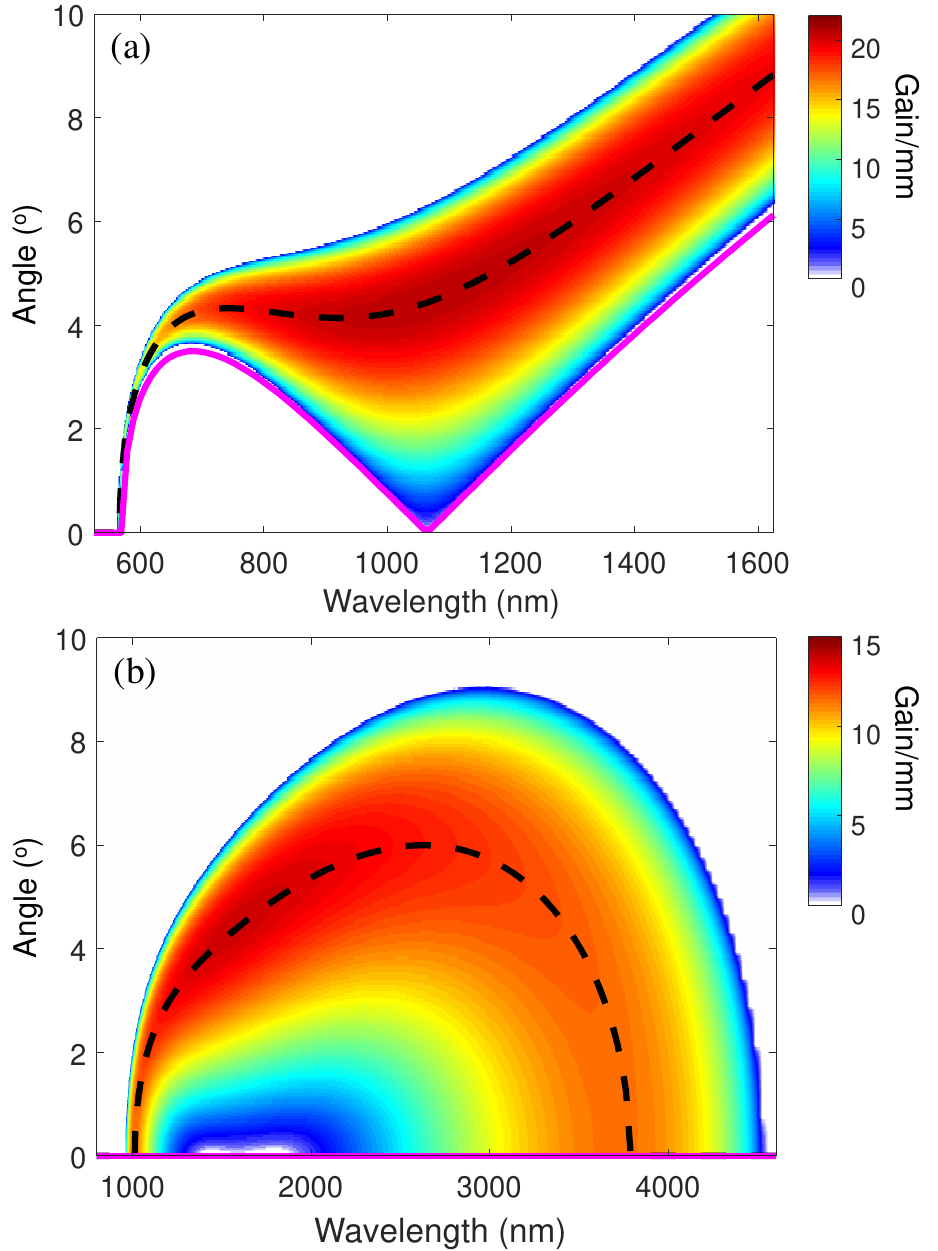}
\caption{KIA with longer wavelength pumps at peak intensity $I_0 = 1.5 \times 10^{13}$ W/cm$^2$ in MgO. (a) Pump central wavelength is 1064 nm; near 4$^\circ$, the gain bandwidth spans more than an octave from 600 to 1400 nm, enabling sub-two-cycle pulses in the visible and near IR. (b) Pump central wavelength at 1600 nm. The longer wavelength pump leads to less gain, but near 4$^\circ$ the bandwidth limited pulse duration is single-cycle in the IR.}
\end{figure}

Although we have shown that we can obtain broadband amplification from a Ti:Sapphire pump, we further study the potential broadband pulse amplification in KIA from other pump sources, shown in Fig. 5. The calculations were again for a peak intensity of $I_0 = 1.5 \times 10^{13}$ W/cm$^2$ in MgO. In (a), we show the effect of pumping with a central wavelength of 1064 nm; near 4$^\circ$, the gain bandwidth spans more than an octave from 600 to 1400~nm. The transform limited pulse duration of this amplified spectrum is 5.2~fs, or sub-two-cycle with a central wavelength around 1000~nm. By increasing the pump central wavelength to 1600~nm, KIA theory predicts less gain, but near 4$^\circ$ the bandwidth-limited pulse duration is single-cycle in the IR, as shown in (b).

Broadband supercontinuum pulses in the infrared can be created via FWM by filamentation in ambient air using a $\omega-2\omega$ FWM setup \cite{FujiOL2007, ChengOL2012, FujiApplSci2017}. A drawback of these few-cycle mid-IR pulses is their beam quality \cite{VoroninAPB2014}, and they tend to be limited to few \textmu J \cite{ThebergePRA2010}. Because KIA produces well-structured near-Gaussian beam shapes, it may be a route to producing single-cycle mid-IR pulses, which are useful for condensed matter high harmonic generation and attosecond science \cite{ShiraiOL2018}.

\section{Conclusions}

In conclusion, we have observed amplification of a supercontinuum spectrum from 500 nm to over 1 \textmu m. The amplification bandwidth depends on the relative pump-seed angle, in agreement with theory. Although Kerr instability amplification theory predicts gain on the order of 25/mm, we only observed amplification of two orders of magnitude, limited by geometrical effects and saturation. We observed that the amplification process did not significantly change the temporal nature of the seed; we expect that chirping the seed could allow for further amplification. Pumping MgO with IR sources can also lead to a broader amplified bandwidths, where single-cycle mid-IR pulses could be used for strong-field condensed matter physics such as high harmonic generation and attosecond science.

\begin{acknowledgments}
SG acknowledges funding from Mitacs Globalink Research Internship. We acknowledge funding from Natural Sciences and Engineering Research Council of Canada (RGPIN-2019-06877) and the University of Windsor Xcellerate grant (5218522). TJH thanks Giulio Vampa, Thomas Brabec, and Claire Duncan for useful conversations.
\end{acknowledgments}

\bibliography{KIA_theory_refs1}

\end{document}